\documentclass[twocolumn]{jpsj3}
\usepackage[final]{changes}

\usepackage{txfonts}
\usepackage{bm}

\title{Field-Induced Superconductivity near the Superconducting Critical Pressure in UTe$_2$}

\author{
Dai~Aoki$^{1,2}$\thanks{E-mail: aoki@imr.tohoku.ac.jp}, 
Motoi~Kimata$^1$,
Yoshiki~J.~Sato$^1$,
Georg~Knebel$^2$,
Fuminori~Honda$^1$,
Ai~Nakamura$^1$,
Dexin~Li$^1$,
Yoshiya~Homma$^1$
Yusei~Shimizu$^1$,
William~Knafo$^3$,
Daniel~Braithwaite$^2$, 
Michal~Vali\v{s}ka$^{2,4}$,
Alexandre~Pourret$^2$,
Jean-Pascal~Brison$^2$, and
Jacques~Flouquet$^2$
}

\inst{%
$^1$IMR, Tohoku University, Oarai, Ibaraki, 311-1313, Japan\\
$^2$Univ. Grenoble Alpes, CEA, Grenoble INP, IRIG, PHELIQS, F-38000 Grenoble, France\\
$^3$Laboratoire National des Champs Magn\'{e}tiques Intenses, UPR 3228, CNRS-UPS-INSA-UGA, 143 Avenue de Rangueil, 31400 Toulouse, France\\
$^4$Charles University, Faculty of Mathematics and Physics, Department of Condensed Matter Physics, Ke Karlovu 5, Prague 2, 121 16, Czech Republic
}

\abst{%
We report the magnetoresistance in the novel spin-triplet superconductor UTe$_2$ under pressure close to the critical pressure $P_{\rm c}$, where the superconducting phase terminates, for field along the three $a$, $b$ and $c$-axes in the orthorhombic structure. 
The superconducting phase for $H\parallel a$-axis just below $P_{\rm c}$ shows a field-reentrant behavior due to the competition with the emergence of magnetic order at low fields. 
The upper critical field $H_{\rm c2}$ for $H\parallel c$-axis shows a quasi-vertical increase in the $H$-$T$ phase diagram just below $P_{\rm c}$, indicating that superconductivity is reinforced by the strong fluctuations which persist even at high fields above $20\,{\rm T}$.
Increasing pressure leads to the disappearance of  superconductivity at zero field with the emergence of magnetic order.
Surprisingly, field-induced superconductivity is observed at high fields, where a spin-polarized state is realized due to the suppression of the magnetic ordered phases;
the spin-polarized state is favorable for superconductivity, whereas the magnetic ordered phase at low field seems to be unfavorable.
The huge $H_{\rm c2}$ in the spin-polarized state seems to imply a spin-triplet state. 
Contrary to the $a$- and $c$-axes, no field-reinforcement of superconductivity occurs for magnetic field along the $b$-axis.
We compare the results with the field-reentrant superconductivity above the metamagnetic field, $H_{\rm m}$ for the field direction tilted by $\sim 30\,{\rm deg.}$ from $b$ to $c$-axis at ambient pressure as well as the field-reentrant (-reinforced) superconductivity in ferromagnetic superconductors, URhGe and UCoGe.
}

\begin{document}
\maketitle
\section{Introduction}
The recent discovery of superconductivity in the heavy fermion paramagnet UTe$_2$ has attracted a lot of attention~\cite{Ran19,Aok19_UTe2}
Although UTe$_2$ at ambient pressure does not show a long-range magnetic order down to low temperatures, 
many similarities to ferromagnetic superconductivity are pointed out.
The microscopic coexistence of ferromagnetism and superconductivity is established in uranium compounds, namely UGe$_2$, URhGe and UCoGe.~\cite{Sax00,Aok01,Huy07},
where the 5$f$ electrons are responsible for the ferromagnetic moments as well as the conduction bands.~\cite{Aok12_JPSJ_review,Aok19} 
A strong Ising-type magnetic anisotropy exists in these systems, 
although the ordered magnetic moment is much smaller than the expected values for a free ion, implying itinerant 5$f$ electrons. 
One of the highlights in ferromagnetic superconductivity is the huge upper critical field, $H_{\rm c2}$.
When the field is applied along the hard-magnetization axis ($b$-axis) in URhGe and UCoGe,
field-reentrant or field-reinforced superconductivity is observed, associated with the collapse of the ferromagnetic Curie temperature, $T_{\rm Curie}$.~\cite{Lev05,Aok09_UCoGe}
The ferromagnetic fluctuations for the field along the hard-magnetization axis are strongly enhanced with the suppression of $T_{\rm Curie}$.
Consequently, the pairing interaction increases, in favor of superconductivity. 
Fermi surface instabilities, such as a Lifshitz transition, at high fields may also be favorable for the field-reinforced superconductivity.~\cite{Bas16,Gou16}

On the other hand, UTe$_2$ with the body-centered orthorhombic structure (space group: {\#}71, $Immm$, $D^{25}_{2h}$) is a paramagnet at ambient pressure.
Superconductivity appears below $T_{\rm c}=1.6\,{\rm K}$.
The Sommerfeld coefficient is $\gamma = 120\,{\rm mJ\,K^{-2}mol^{-1}}$, indicating the heavy electronic state. 
The large specific heat jump at $T_{\rm c}$ displays strong coupling superconductivity.
The large residual $\gamma$-value even in the high quality samples is one of the enigma, suggesting a partially gapped superconductivity, such as the $A_1$ state in superfluid $^3$He~\cite{Mac20}. 
However, it should be stressed that the direct 2$^{\rm nd}$ order transition from the paramagnetic state to a non-unitary state as the $A_1$ state is forbidden, according to \deleted{the} symmetry arguments.~\cite{Min08}
Thus, the possibility of a double transition at $T_{\rm c}$, or the development of short range magnetic order at higher temperature are \deleted[id=JF]{discussed from the experimental results, which is} still under debate.~\cite{Hay20,Kit20,Cai20}

The most remarkable feature of UTe$_2$ is the huge $H_{\rm c2}$ and the field-reentrant superconductivity.~\cite{Kne19,Ran19_HighField}
For $H\parallel b$-axis (hard-magnetization axis), field-reentrant superconductivity is observed above $16\,{\rm T}$ in the $H$-$T$ phase diagram,
and it abruptly disappears above the metamagnetic field $H_{\rm m}=35\,{\rm T}$,
at which the first order transition with a sharp jump of magnetization and the associated mass enhancement occur.~\cite{Miy19,Kna19,Ima19}
The upper critical field $H_{\rm c2}(0)$ for $a$ and $c$-axes also shows large values ($7$ and $11\,{\rm T}$, respectively), exceeding the Pauli limit for all field directions, supporting a spin-triplet scenario, which is evidenced by the spin susceptibility in NMR experiments.~\cite{Nak19}

\begin{figure}[tbh]
\begin{center}
\includegraphics[width=\hsize,clip]{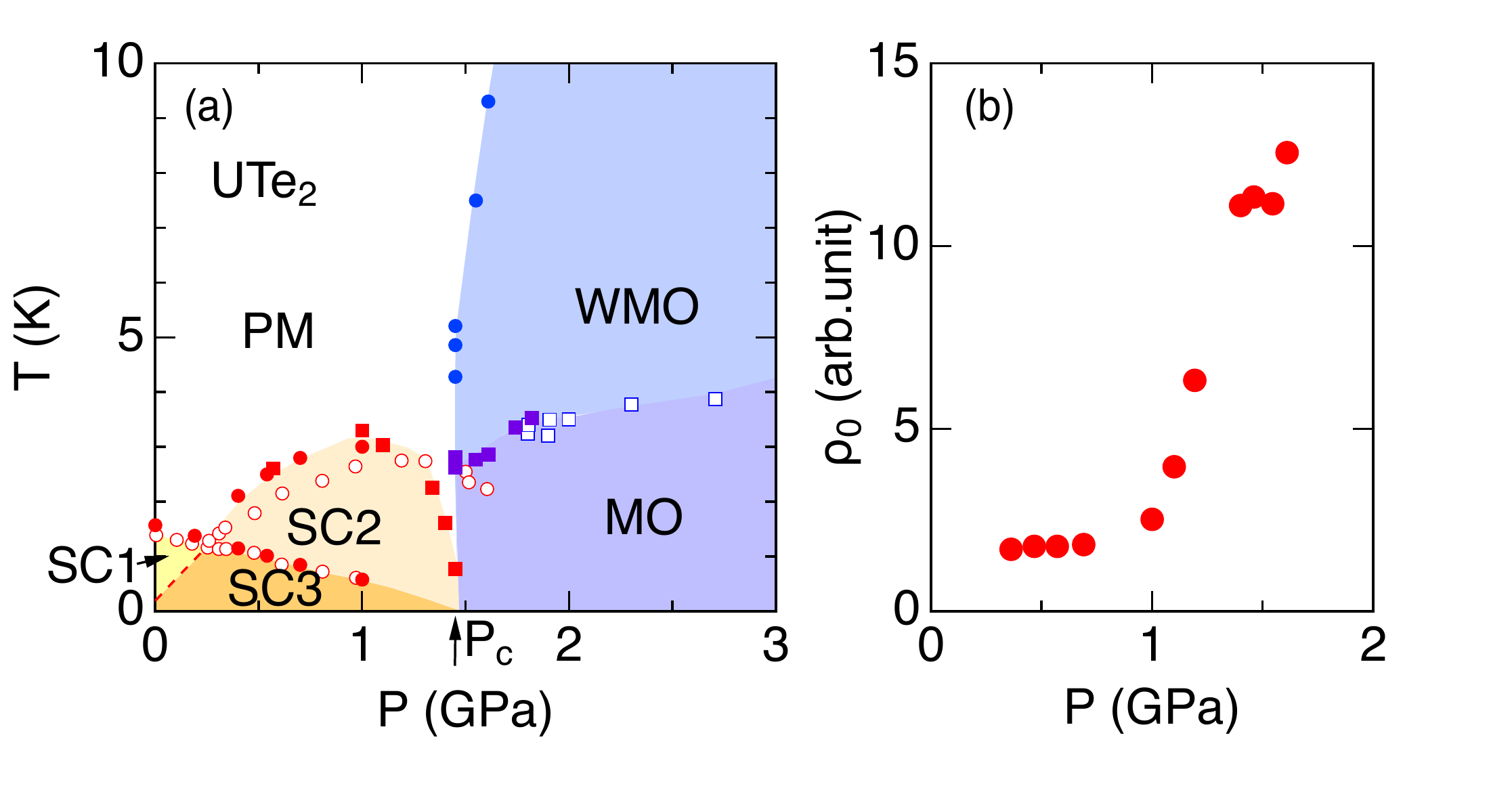}
\end{center}
\caption{(Color online) (a) $T$-$P$ phase diagram at zero field in UTe$_2$. SC1, SC2 and SC3 denote the multiple superconducting phases. PM, MO and WMO denote paramagnetism and magnetic ordered phase and weakly magnetic order, respectively. Here we define $P_{\rm c}\sim 1.45\,{\rm GPa}$ as the critical pressure of suppression for superconducting phases. The data denoted by open circles and open squares are cited from Ref.~\citen{Bra19}.
(b) Pressure dependence of the residual resistivity at zero field.}
\label{fig:TP_phase}
\end{figure}
Another important point in UTe$_2$ is the emergence of multiple superconducting phases under pressure ($P$) clarified by AC calorimetry measurements.~\cite{Bra19}
As shown in Fig.~\ref{fig:TP_phase}(a), 
$T_{\rm c}$ initially decreases with pressure and splits into two transitions above $0.25\,{\rm GPa}$.
The lower $T_{\rm c}$ continuously decreases with further increasing pressure \added{and extrapolates linearly to zero at the critical pressure $P_{\rm c} \sim 1.45\,{\rm GPa}$},  
whereas the higher $T_{\rm c}$ increases, showing the maximum at $3\,{\rm K}$ at $P \sim 1\,{\rm GPa}$,
and decreases rapidly, on approaching $P_{\rm c}$.
Superconductivity is suppressed at the critical pressure $P_{\rm c}\sim 1.45\,{\rm GPa}$, and new long range magnetically ordered phases appear.~\cite{Bra19, Ran20_pressure, Aok20_UTe2}
The transition to the magnetic state is accompanied by a tiny valence instability.~\cite{Tho20} 
Right now, the microscopic nature of the order is still unclear, but there are indications that an antiferromagnetic state forms.~\cite{Aok20_UTe2,Tho20}.

AC calorimetry measurements under pressure with the magnetic field applied along the $a$-axis confirmed the occurrence of multiple superconducting phases in UTe$_2$ and that the sudden increase of $H_{\rm c2}$ at low temperatures is connected with the lower $T_{\rm c}$.~\cite{Aok20_UTe2}
These experimentally observed superconducting phases imply different order parameters.~\cite{Mac20,Shi20,Ish20}
This is also consistent with a spin-triplet state, which could have multiple phases owing to the spin degree of freedom.

The remarkable evolution of $H_{\rm c2}(T)$ under pressure was clarified for $H\parallel a$, $b$ and $c$-axes~\cite{Kne20}.
This is, of course, related to the interplay between superconductivity and the magnetic order which appears above $\sim P_{\rm c}$.
In particular, the initial slope of $H_{\rm c2}$ for $H\parallel c$-axis at low fields near $P_{\rm c}$ is exceptionally high.~\cite{Bra19,Kne20}

Here, we present magnetoresistance measurements under pressure, focusing on the results near $P_{\rm c}$ for $H\parallel a$, $b$ and $c$-axes.
Just above $P_{\rm c}$, field-induced superconductivity was found at high field at least up to $27\,{\rm T}$ in the spin-polarized state for $H\parallel c$-axis.
Further slightly increasing pressure, this field-induced superconducting phase is pushed up to higher field and lower temperature region, corresponding to the evolution of the magnetic phase.
The results indicate that superconductivity can survive in the spin-polarized state far above the critical field of the magnetic ordered state,
while magnetic ordered state seems to suppress superconductivity. 

\section{Experimental}
High quality single crystals of UTe$_2$ were grown using the chemical vapor transport method.
The off-stoichiometric amounts of starting materials with the atomic ratio U : Te = 1 : 1.5 were put into 
the quartz ampoule, which was sealed under vacuum together with iodine as the transport agent. 
The quartz ampoule was slowly heated up to $900\,^\circ{\rm C}$ and maintained for pre-reaction. 
Then temperature gradient $1060/1000\,^\circ{\rm C}$ was applied for two weeks.
At lower temperature side, many single crystals of typically a few millimeters in length were obtained.
Single crystals were checked by single crystal X-ray analysis, 
revealing that the crystallographic parameters are in good agreement with the previously reported values.~\cite{Ike06_UTe2}
We also carefully checked the stoichiometry of U and Te, using scanning electron microprobe (SEM) and ICP spectroscopy.
We do not detect a large deviation from the stoichiometric ratio $\mbox{Te/U}\sim 2$.

The magnetoresistance was measured by the four-probe AC or DC method in a piston cylinder cell at pressure up to $1.6\,{\rm GPa}$ with Daphne 7373 as pressure transmitting medium.
Three samples for $H\parallel a$, $b$ and $c$-axes were put together in the same pressure cell.
The electrical current was applied along the $a$-axis for all samples. 
The pressure was determined by the superconducting transition temperature of lead. 
As we use different crystals and small pressure inhomogeneity, the critical pressure for the three samples are not exactly the same.
The pressure cell attached with a local thermometer was inserted in a top-loading dilution refrigerator offering a minimum temperature of $30\,{\rm mK}$ and magnetic fields up to $15\,{\rm T}$ using a superconducting magnet. 
In addition, for higher field up to $30\,{\rm T}$, a hybrid magnet at the high field laboratory in Sendai was used together with a $^3$He or $^4$He cryostat.

\section{Results and Discussion}
Figure~\ref{fig:TP_phase}(a) shows the $T$-$P$ phase diagram at zero field.
The data points are plotted from the present results as well as previous results by resistivity and AC calorimetry measurements.~\cite{Bra19,Aok20_UTe2,Kne20}
Similar phase diagrams are reported in Refs.~\citen{Ran20_pressure,Tho20}.
As we already mentioned, multiple superconducting phases appear under pressure,
and superconductivity collapses at $P_{\rm c}\sim 1.45\,{\rm GPa}$. 
At $P\gtrsim P_{\rm c}$, a magnetic order (MO), which is most likely antiferromagnetic rather than ferromagnetic, appears around $3\,{\rm K}$,
which gradually increases with pressure (see also Ref.~\citen{Tho20}). 
In addition, a broad bump occurs in the resistivity at higher temperature together with a broad anomaly in the specific heat, as discussed below.
With pressure, the temperature for this anomaly increases rapidly with pressure (see also Refs.~\citen{Ran20_pressure,Tho20}).
The microscopic nature of this high temperature weak magnetic ordered (WMO) phase, which is not driven by the long-range magnetic order, is also not clarified yet. 

In resistivity measurements, a drastic increase of residual resistivity, $\rho_0$ is observed as shown in Fig.~\ref{fig:TP_phase}(b), which is also inferred from Ref.~\citen{Kne20}.
This indicates a drastic change of the electronic state associated with a tiny change of the U-valence at $P_{\rm c}$, which leads to a Fermi surface reconstruction. 
Since the bare LDA band structure calculation predicts a Kondo semiconducting nature with a narrow gap at the Fermi energy with flat bands,~\cite{Aok19_UTe2,Har20} 
the Fermi surface can be easily modified by external parameters, such as pressure,
due to changes in nature of the electronic and magnetic correlations.
This is also supported by LDA+$U$ calculation, which shows that the Fermi surface is very sensitive to the Coulomb repulsion, $U$, and an insulator to metal transition can be induced by varying $U$.~\cite{Ish19}

\begin{figure}[tbh]
\begin{center}
\includegraphics[width= \hsize,clip]{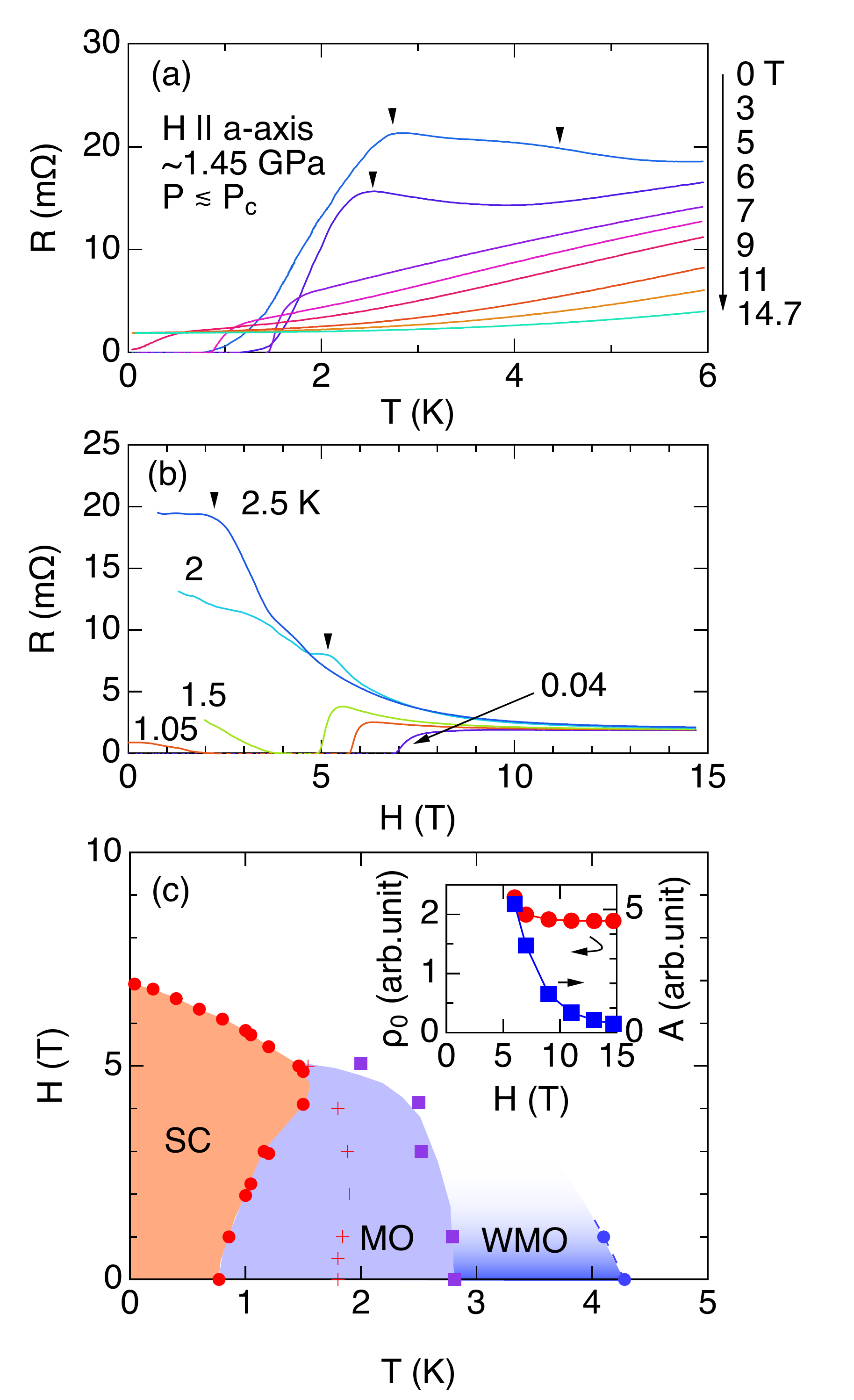}
\end{center}
\caption{(Color online) (a) Temperature dependence of the magnetoresistance at different fields for $H\parallel a$-axis at around $1.45\,{\rm GPa}$, just below the superconducting critical pressure $P_{\rm c}$.
(b) Field dependence of the magnetoresistance at different temperatures.
(c) $H$-$T$ phase diagram for $H\parallel a$-axis just below $P_{\rm c}$.
Red crosses indicate the midpoint of the resistivity drop due to the inhomogeneous superconductivity.
The inset shows the field dependence of the residual resistivity $\rho_0$ and the $A$ coefficient extracted from the fitting 
with $\rho = \rho_0 + AT^2$.
}
\label{fig:a_axis}
\end{figure}
Next we focus on the results for $H \parallel a$-axis at $\sim 1.45\,{\rm GPa}$ close to $P_{\rm c}$.
The temperature dependence of the magnetoresistance at different fields is shown in Fig.~\ref{fig:a_axis}(a). 
On cooling, at zero field, a first anomaly is observed near $4.3\,{\rm K}$, while a second anomaly appears at $2.8\,{\rm K}$.
Under magnetic field, both anomalies shift to lower temperatures, disappearing above $5\,{\rm T}$.
Superconductivity defined by $\rho =0$ is observed at $T_{\rm c}\sim 0.7\,{\rm K}$ after gradual decrease below $2.8\,{\rm K}$ at zero field.
Interestingly, $T_{\rm c}$ initially increases with field and then decreases above $5\,{\rm T}$,
revealing the field-reinforced superconductivity.

In the field scan, the field-reinforced superconductivity is more significant.
As shown in Fig.~\ref{fig:a_axis}(b), superconductivity appears from $0$ to $7\,{\rm T}$ at $0.04\,{\rm K}$,
but at higher temperature, $1.05\,{\rm K}$, superconductivity is observed at finite field range from $2$ to $5.8\,{\rm T}$.
At $2$ and $2.5\,{\rm K}$, superconductivity is no more visible, but a magnetic anomaly is detected.
It should be noted that the magnetoresistance at high temperature, $5\,{\rm K}$, decreases significantly with field.
This is also shown in the field dependence of the residual resistivity, $\rho_0$ (see the inset of Fig.~\ref{fig:a_axis}(c)), which is extracted from the $T^2$-dependence of resistivity, namely $\rho = \rho_0 +AT^2$.
The rapid decrease of $A$ with field suggests that magnetic fluctuations are suppressed at high fields.

Figure~\ref{fig:a_axis}(c) shows the $H$-$T$ phase diagram for $H\parallel a$-axis just below $P_{\rm c}$.
Superconductivity is reinforced under magnetic field, revealing the maximum of $T_{\rm c}$ at $1.5\,{\rm K}$ around $5\,{\rm T}$, which is linked to the collapse of MO. 
$H_{\rm c2}$ further increases with field and reaches $7\,{\rm T}$ at $0\,{\rm K}$.
The determination of the phase line indicating the collapse of the magnetic phase needs further thermodynamic experiments. 
However, taking into account the reduced $T_{\rm c}$ at zero field,
superconductivity is obviously in competition with the occurrence of long range magnetic order.
This is also demonstrated by the results for $H\parallel c$-axis as shown later.
We also remark that the $H_{\rm c2}(T)$ curve at lower pressure between $0.5\,{\rm GPa}$ and $1\,{\rm GPa}$
shows a strong convex curvature near $T_{\rm c}$, resembling the Pauli paramagnetic effect.~\cite{Kne20,Aok20_UTe2} 
The strong convex curvature may indicate a precursor for the magnetic order at higher pressure.

\begin{figure}[tbh]
\begin{center}
\includegraphics[width= 0.9\hsize,clip]{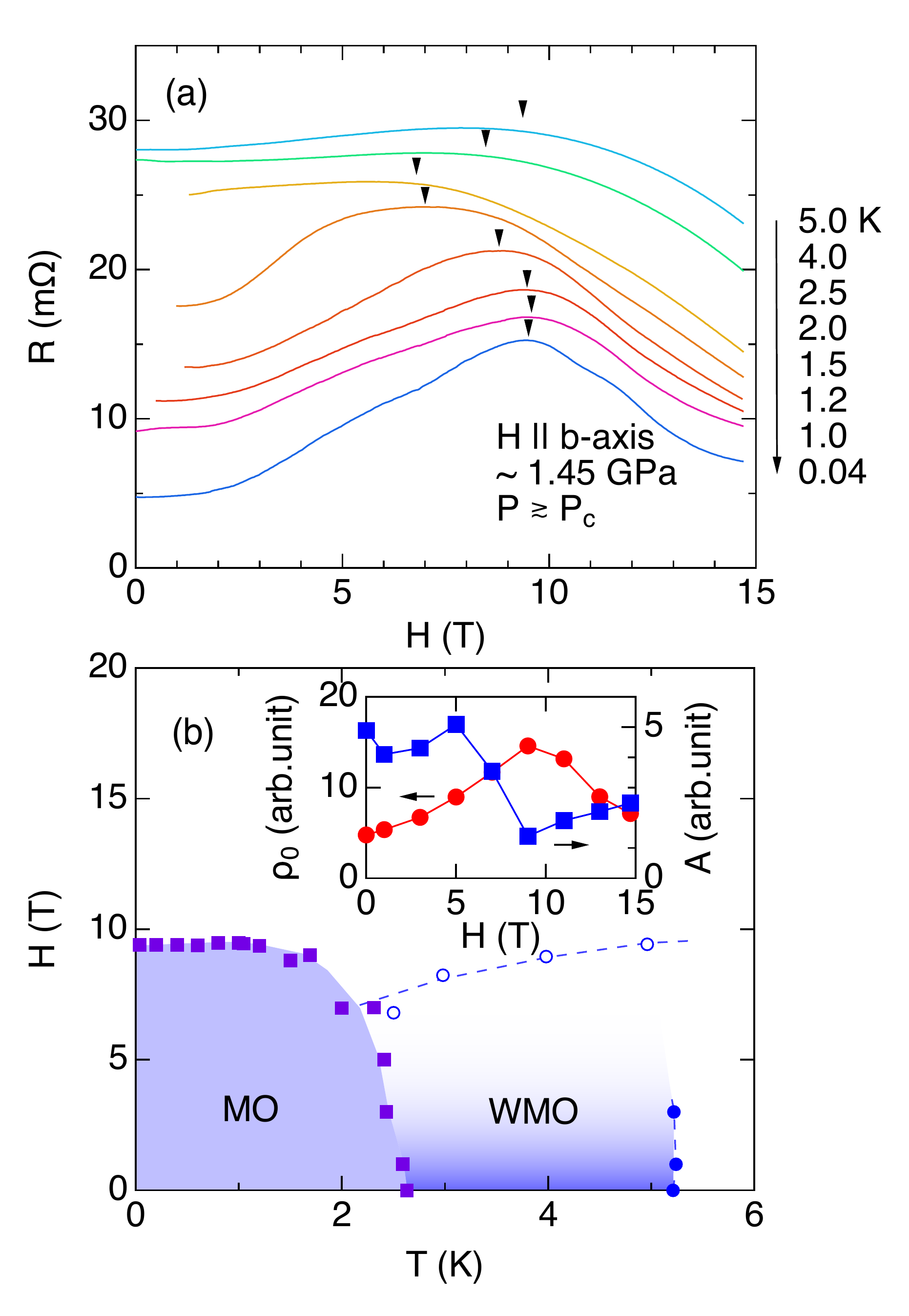}
\end{center}
\caption{(Color online) (a) Field dependence of the magnetoresistance at different temperatures for $H\parallel b$-axis at around $1.45\,{\rm GPa}$ just above $P_{\rm c}$.
(b) $H$-$T$ phase diagram for $H\parallel b$-axis just above $P_{\rm c}$. Open circles indicate the crossover obtained from the broad anomaly in magnetoresistance. The inset shows the field dependence of the residual resistivity $\rho_0$ and the $A$ coefficient.}
\label{fig:b_axis}
\end{figure}
Figure~\ref{fig:b_axis}(a) shows the field dependence of the magnetoresistance for $H\parallel b$-axis just above the critical pressure $P_{\rm c}$.
No superconductivity is detected down to the lowest temperature, $0.04\,{\rm K}$.
At the lowest temperature, a broad maximum is observed around $9.5\,{\rm T}$, which is reduced to $\sim 7\,{\rm T}$ at $2\,{\rm K}$.
This anomaly is connected to the magnetic order at $2.5\,{\rm K}$ at zero field, as shown in the $H$-$T$ phase diagram (see Fig.~\ref{fig:b_axis}(b)). 
Kinks in the field dependence of the magnetoresistance for temperatures below $2\,{\rm K}$ may be a signature of slight changes in the magnetic structure, and present another indication of an antiferromagnetic order.
At a temperature higher than $2.5\,{\rm K}$, a broad anomaly is still visible and shifts to the higher field, indicating a crossover line.
In the temperature scan (not shown in the figure), the higher temperature anomaly around $5\,{\rm K}$ is visible up to $3\,{\rm T}$,
and disappears at higher fields.
Contrary to the case for $H\parallel a$ or $c$-axis (see  below), 
field-reinforced (-reentrant) superconductivity is not observed near $P_{\rm c}$ for $H\parallel b$-axis.
Below $P_{\rm c}$, the superconducting phase is cut off above $H_{\rm m}$ due to the first order metamagnetic transition associated with a Fermi surface reconstruction,
which is reduced from $H_{\rm m}\sim 35\,{\rm T}$ at ambient pressure to $H_{\rm m}\sim 5\,{\rm T}$ at $1.4\,{\rm GPa}$.~\cite{Kne20}

\begin{figure}[tbh]
\begin{center}
\includegraphics[width= 0.8\hsize,clip]{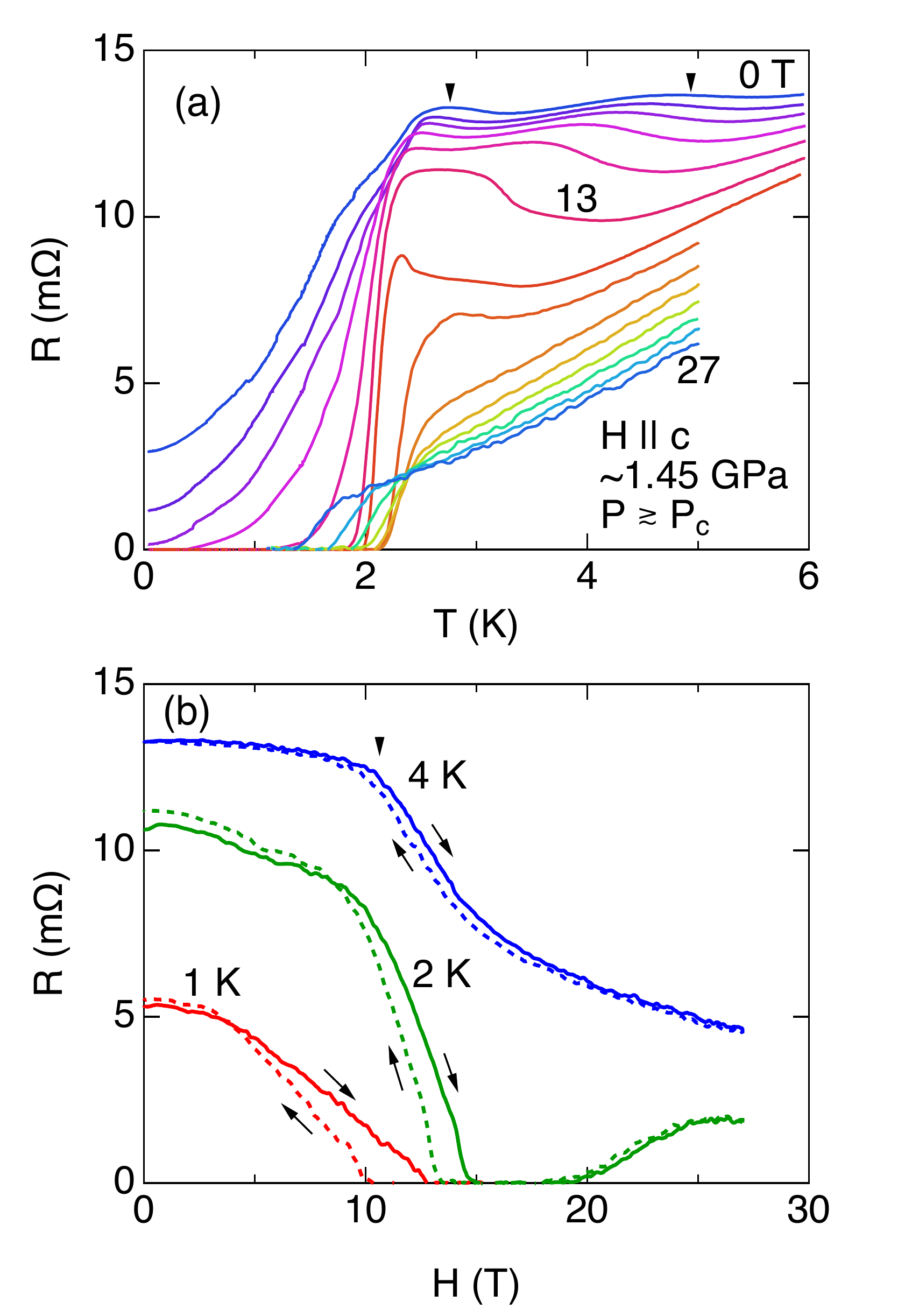}
\end{center}
\caption{(Color online) (a) Temperature dependence of the magnetoresistance for $H\parallel c$-axis just above $P_{\rm c}$
at $0$, $5$, $7$, $9$, $11$, $13$, $14.7$, $15$, $17$, $19$, $21$, $23$, $25$ and $27\,{\rm T}$.
(b) Field dependence of the magnetoresistance at different temperatures. The solid and dotted lines are up and down-sweeps of field, respectively.}
\label{fig:c_axis_MR}
\end{figure}
Next we show in Fig.~\ref{fig:c_axis_MR} the temperature and field dependence of magnetoresistance for $H\parallel c$-axis at $P = 1.45$~GPa, just above $P_{\rm c}$.
At zero field, clearly two magnetic anomalies are observed on cooling at  $2.8$ and $5\,{\rm K}$. 
Below $2.8\,{\rm K}$ the magnetoresistance decreases, but below $1.8\,{\rm K}$ an even stronger decrease occurs. 
We define this kink as the onset of superconductivity, 
but $\rho = 0$ is not observed down to the lowest temperature of $0.04\,{\rm K}$.
The lower temperature anomaly is slightly reduced with increasing field, 
whereas the higher temperature anomaly becomes sharp and is rapidly reduced with field. 
The two magnetic anomalies merge and are invisible anymore above $\sim 15\,{\rm T}$.
Interestingly, $\rho = 0$ is observed in the temperature scan at $H \gtrsim 9\,{\rm T}$, and at higher fields the superconducting transition becomes sharper, revealing field-induced superconductivity.
Superconductivity is observed up to our highest field, $27\,{\rm T}$ with slightly reduced $T_{\rm c}$.
The temperature dependence at high field follows a $T^2$-dependence with Fermi liquid behavior in the normal state for $H > 15\,{\rm T}$.
In the field scan, the field-induced superconductivity is more significant, as shown in Fig.~\ref{fig:c_axis_MR}(b).
At $1\,{\rm K}$, superconductivity appears above $13\,{\rm T}$ in the up-sweep field.
In the down sweep, superconductivity disappears below $10\,{\rm T}$, showing a clear hysteresis due to the first order transition.
At $2\,{\rm K}$, superconductivity is observed only in the limited field range between $\sim 14\,{\rm T}$ and $20\,{\rm T}$.
At $4\,{\rm K}$, no superconductivity is detected anymore, while the magnetic anomaly is detected at $11\,{\rm T}$ with a small hysteresis.
Note that a similar field-induced superconductivity is reported in Ref.~\citen{Ran20_pressure}, in which the field direction was not clearly indicated.

\begin{figure}[tbh]
\begin{center}
\includegraphics[width= 0.8\hsize,clip]{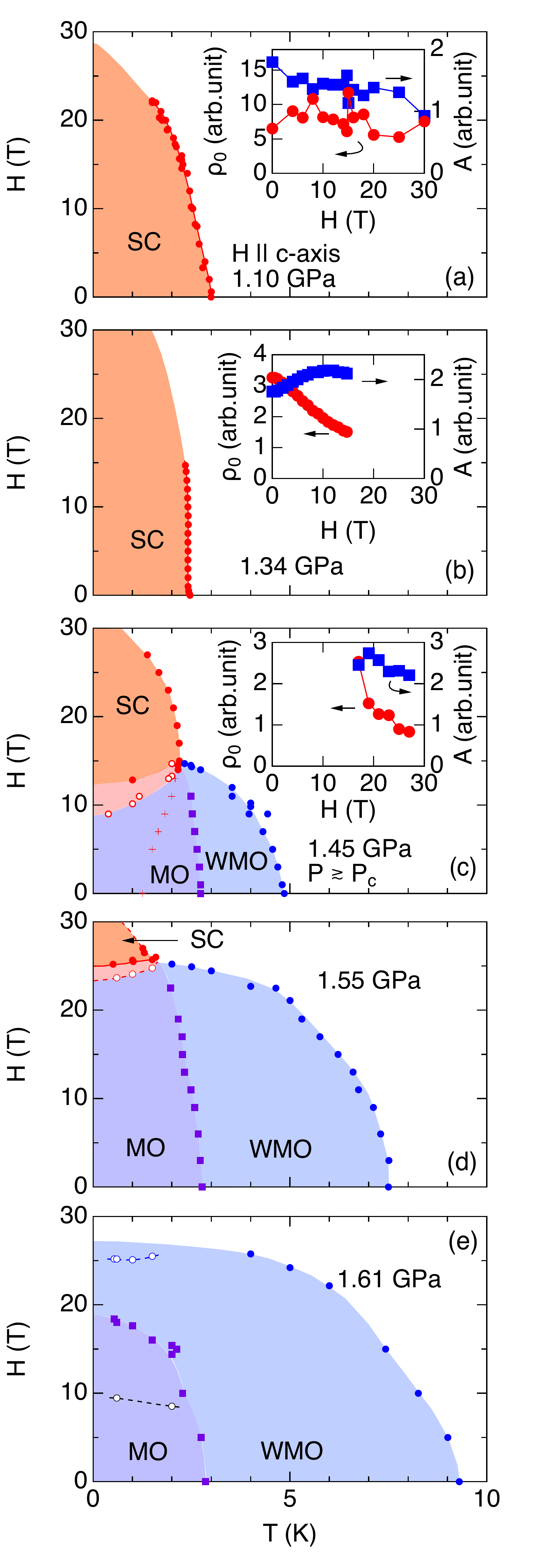}
\end{center}
\caption{(Color online) $H$-$T$ phase diagrams for $H\parallel c$-axis at $1.10$, $1.34$, $\sim 1.45\gtrsim P_{\rm c}$, $1.55$ and $1.61\,{\rm GPa}$.
Red crosses in panel (c) indicate the midpoint of the resistivity drop due to the inhomogeneous superconductivity.
The insets show the field dependence of the residual resistivity and the $A$ coefficient at different pressures.}
\label{fig:c_axis}
\end{figure}
The $H$-$T$ phase diagrams for $H\parallel c$-axis near $P_{\rm c}$ are summarized in Fig.~\ref{fig:c_axis}.
At $1.1\,{\rm GPa}$, $T_{\rm c}$ reveals the maximum, $\sim 3\,{\rm K}$ in the $T$-$P$ phase diagram, as already shown in Fig.~\ref{fig:TP_phase}(a).
The initial slope of $H_{\rm c2}$, $|dH_{\rm c2}/dT|_{H=0} \approx 20\,{\rm T/K}$ is very large (see Fig.~\ref{fig:c_axis}(a)),
thus $H_{\rm c2}$ at low temperature is very large, surviving at least up to $27\,{\rm T}$.
The residual resistivity, $\rho_0$ is nearly unchanged with field, while the resistivity coefficient, $A$, gradually decreases,
as shown in the inset of Fig.~\ref{fig:TP_phase}(a).
At $1.34\,{\rm GPa}$, the initial slope is almost vertical as shown in Fig.~\ref{fig:c_axis}(b).
$H_{\rm c2}$ is nearly unchanged up to the maximum measured field here, $14.7\,{\rm T}$.
The $A$ coefficient shows a broad maximum, retaining the large value, while $\rho_0$ rapidly decreases with field, suggesting a precursor of the magnetic order at higher pressure. 
At the pressure just above $P_{\rm c}\sim 1.45\,{\rm GPa}$, the magnetic order appears, and superconductivity is not detected at zero field, as shown in Fig.~\ref{fig:c_axis}(c).
When the magnetic ordered phase is suppressed with field, the field-induced superconductivity is observed in the spin-polarized state.
Correspondingly, $\rho_0$ decreases rapidly above $17\,{\rm T}$, probably due to the spin polarization.
However, the $A$ coefficient does not show a large decrease, indicating remaining fluctuations even in the spin-polarized state,
which contrasts wtih the remarkable suppression of $A$ for $H\parallel a$-axis as shown in the inset of Fig.~\ref{fig:a_axis}(c).
On further increasing pressure, the superconducting phase is pushed away to the high-field and low-temperature region, corresponding to the development of MO and WMO, as shown in Fig.~\ref{fig:c_axis}(d).
Finally at $1.61\,{\rm GPa}$, superconductivity disappears. 
In addition, other anomalies are detected in the magnetic ordered phases through the field scan, probably due to the change of magnetic and/or electronic structure.

The evolution of $H$-$T$ phase diagrams in Fig.~\ref{fig:c_axis} indicates that the emergence of magnetism suppresses superconductivity, which survives for $P > P_{\rm c}$ only in the spin polarized state.
This situation is similar to the field-reentrant superconductivity observed above $H_{\rm m}$ for the field direction titled by $\sim 30\,{\rm deg}$ from $b$ to $c$-axis,
where the spin-polarized state is realized above the metamagnetic field $H_{\rm m}$ with a sharp jump of magnetization.~\cite{Ran19_HighField,Kna21}

Initially it has been proposed that superconductivity in UTe$_2$ occurs at the border of a ferromagnetic state and that the concomitant ferromagnetic fluctuations are responsible for the spin-triplet superconductivity.~\cite{Sun19,Tok19}
Contrarily, the recent high pressure studies raise questions about the pure ferromagnetic state under pressure, 
and there are strong evidences for antiferromagnetic order rather than ferromagnetic order above $P_{\rm c}$.~\cite{Aok20_UTe2,Kne20,Tho20}
The lower transition for MO has clearly the characteristics of an antiferromagnetic ordered state. 
Remarkably, the pressure induced transition from the superconducting to the magnetic ordered state seems to be of first order
and there may be a finite pressure regime where both orders coexists inhomogeneously. 
This is clearly evidenced by the rather large and incomplete superconducting transitions which occur inside the magnetically ordered state. This inhomogeneous coexistence of superconductivity and magnetism is very similar to that observed close to the critical point in the antiferromagnets CeRhIn$_5$, CeIrSi$_3$ and CeRhSi$_3$.~\cite{Kne06,Par06,Set11,Kim07} 
The main difference is that due to the extremely large upper critical field in UTe$_2$ superconductivity survives far above the magnetically ordered state, while in the antiferromagnets (CeRhIn$_5$, CeIrSi$_3$, CeRhSi$_3$) the magnetic state is favorable under magnetic field over the superconducting state.

The competition between ferromagnetism, antiferromagnetism and sharp crossover between weakly and strongly polarized state 
is well illustrated in the case of CeRu$_2$Ge$_2$~\cite{Wil04} and CeRu$_2$Si$_2$ under pressure or doping~\cite{Fis91,Aok12_CeRu2Si2}.
Similar interplays may occur here in UTe$_2$. 
However, it must be stressed that the novelty in U heavy fermion materials is that ``hidden'' valence fluctuations may exist between U$^{3+}$ and U$^{4+}$ configurations which can both lead to long range magnetic order. 
The key point under pressure is that the renormalization of one of the configurations will be associated with a drastic change in the crystal field.
Consequently, there may be a strong feedback on the relative magnetic anisotropy between three axes.
The extra novelty is that the band structure itself is directly linked to the nature and strength of the correlations.
At first glance, the interplay of ferromagnetic and antiferromagnetic fluctuations appear as a main ingredient to stabilize unique superconducting phases, 
as both antiferromagnetic and ferromagnetic fluctuations may exist as it is inferred from theories~\cite{Ish20,Xu19} and experiments~\cite{Tok19,Sun19,Dua20}.
A key parameter is the pressure and field variation of the volume and thus of the valence opening an additional source of superconductivity via associated valence fluctuations.\cite{Oni00, Wat06} 
Measurements of magnetostriction must be an important target knowing the strong first order nature of many detected transitions.

Let us compare the present results with the case for ferromagnetic superconductivity.
In UCoGe and URhGe, the field-reentrant or field-reinforced superconductivity is observed 
when the field is applied along the hard-magnetization axis ($b$-axis).
The enhancement of ferromagnetic fluctuations are induced by the suppression of $T_{\rm Curie}$ in the anisotropic Ising system.
These field-induced ferromagnetic fluctuations are a main driving force for superconductivity.
At higher field, superconductivity disappears rapidly because ferromagnetic fluctuations become weak.
In UTe$_2$ at ambient pressure, the magnetization jump ($\sim 0.5\,\mu_{\rm B}$) at $H_{\rm m}$ is five times larger than that observed at the spin-reorientation field $H_{\rm R}$ in URhGe. 
Above $H_{\rm m}$, the $b$-axis becomes the easy-magnetization axis in UTe$_2$, 
while in URhGe the $b$-axis will be the easy-magnetization axis only above $H\sim 18\,{\rm T}$~\cite{Har11}, which is higher than $H_{\rm R}\sim 12\,{\rm T}$. 
In UCoGe, the $c$-axis remains the easy ones at least up to $60\,{\rm T}$~\cite{Kna12};
no metamagnetism for switching easy-axis occurs, which may be coupled to $H_{\rm c2}(0)\sim 25\,{\rm T}$ for $a$-axis. 
Contrary to URhGe and UCoGe, the huge jump of magnetization in UTe$_2$ will drive a strong volume change, leading to a feedback for valence instabilities

In UTe$_2$, the occurrence of superconductivity in a spin-polarized state appears quite different.
One novelty is the mark of antiferromagnetic order (MO) coupled with the strong correlations (WMO).
Our observation points out that pressure and magnetic field lead to drastic changes of the superconducting boundaries.
A next important step will be to determine the nature of the MO and WMO phases, the link between the MO, $H_{\rm MO}(T)$ boundary with $H_{\rm c2}(T)$, the pressure and field evolution of the electronic structure, and the nature of the pairing mechanism.
A first issue is to determine a switching mechanism from the huge metamagnetic field $H_{\rm m}\sim 35\,{\rm T}$ for $H\parallel b$-axis at ambient pressure to the magnetic critical field $H_{\rm MO}(0)$,
 which is $6$, $10$, and $15\,{\rm T}$ at $1.45\,{\rm GPa}$ for $H\parallel a$, $b$, and $c$-axes, respectively.

\section{Summary}
The magnetoresistance measurements along the three main $a$, $b$, and $c$-axes in UTe$_2$ gives a rather complex response to the competition of superconductivity and magnetic order near $P_{\rm c}$.
The field-induced superconductivity was observed just above $P_{\rm c}$ for $H\parallel c$-axis at high fields, where the spin-polarized state is realized with the collapse of magnetic ordered phases, revealing the spin-polarized state is favorable for superconductivity.
Next targets will be to determine precisely the pressure evolution of the magnetic parameters such as the magnetic anisotropy, the interplay between antiferromagnetic and ferromagnetic correlations, the volume changes as functions of pressure and field, and of course the corresponding change of the Fermi surface. 
With these knowledges, understanding of the different superconducting phases appeared below $P_{\rm c}$ will be certainly clarified.

\section*{Acknowledgements}
We thank K. Miyake, A. Miyake, H. Harima, K. Machida, Y. Yanase, V. Mineev, S. Fujimoto, K. Ishida, Y. Tokunaga and F. Hardy
for fruitful discussion.
This work was supported by ERC starting grant (NewHeavyFermion), ANR (FRESCO) and KAKENHI (JP19H00646, JP20K20889, JP20H00130, JP20KK0061), GIMRT (20H0406), and ICC-IMR.


\begin{thebibliography}{10}

\bibitem{Ran19}
S.~Ran, C.~Eckberg, Q.-P. Ding, Y.~Furukawa, T.~Metz, S.~R. Saha, I.-L. Liu,
  M.~Zic, H.~Kim, J.~Paglione, and N.~P. Butch: 
  Science {\bfseries 365} 684 (2019).


\bibitem{Aok19_UTe2}
D.~Aoki, A.~Nakamura, F.~Honda, D.~Li, Y.~Homma, Y.~Shimizu, Y.~J. Sato,
  G.~Knebel, J.-P. Brison, A.~Pourret, D.~Braithwaite, G.~Lapertot, Q.~Niu,
  M.~Vali{\v{s}}ka, H.~Harima, and J.~Flouquet: J. Phys. Soc. Jpn. {\bfseries
  88} 043702 (2019).

\bibitem{Sax00}
S.~S. Saxena, P.~Agarwal, K.~Ahilan, F.~M. Grosche, R.~K.~W. Haselwimmer, M.~J.
  Steiner, E.~Pugh, I.~R. Walker, S.~R. Julian, P.~Monthoux, G.~G. Lonzarich,
  A.~Huxley, I.~Sheikin, D.~Braithwaite, and J.~Flouquet: Nature {\bfseries
  406} 587 (2000).

\bibitem{Aok01}
D.~Aoki, A.~Huxley, E.~Ressouche, D.~Braithwaite, J.~Flouquet, J.-P. Brison,
  E.~Lhotel, and C.~Paulsen: Nature {\bfseries 413} 613 (2001).

\bibitem{Huy07}
N.~T. Huy, A.~Gasparini, D.~E. {de Nijs}, Y.~Huang, J.~C.~P. Klaasse,
  T.~Gortenmulder, A.~{de Visser}, A.~Hamann, T.~{G\"{o}rlach}, and
  H.~v.~{L\"{o}hneysen}: Phys. Rev. Lett. {\bfseries 99} 067006 (2007).

\bibitem{Aok12_JPSJ_review}
D.~Aoki and J.~Flouquet: J. Phys. Soc. Jpn. {\bfseries 81} 011003 (2012).

\bibitem{Aok19}
D.~Aoki, K.~Ishida, and J.~Flouquet: J. Phys. Soc. Jpn. {\bfseries 88}  022001 (2019).

\bibitem{Lev05}
F.~L\'{e}vy, I.~Sheikin, B.~Grenier, and A.~D. Huxley: Science {\bfseries 309} 1343
  (2005).

\bibitem{Aok09_UCoGe}
D.~Aoki, T.~D. Matsuda, V.~Taufour, E.~Hassinger, G.~Knebel, and J.~Flouquet:
  J. Phys. Soc. Jpn. {\bfseries 78} 113709 (2009).

\bibitem{Bas16}
G.~Bastien, A.~Gourgout, D.~Aoki, A.~Pourret, I.~Sheikin, G.~Seyfarth,
  J.~Flouquet, and G.~Knebel: Phys. Rev. Lett. {\bfseries 117} 206401 (2016).

\bibitem{Gou16}
A.~Gourgout, A.~Pourret, G.~Knebel, D.~Aoki, G.~Seyfarth, and J.~Flouquet:
  Phys. Rev. Lett. {\bfseries 117} 046401 (2016).

\bibitem{Mac20}
K.~Machida: J. Phys. Soc. Jpn. {\bfseries 89} 033702 (2020).

\bibitem{Min08}
V.~P. Mineev: J. Phys. Soc. Jpn. {\bfseries 77} 103702 (2008).

\bibitem{Hay20}
I.~M. Hayes, D.~S. Wei, T.~Metz, J.~Zhang, Y.~S. Eo, S.~Ran, S.~R. Saha,
  J.~Collini, N.~P. Butch, D.~F. Agterberg, A.~Kapitulnik, and J.~Paglione:
  arXiv:2002.02539 .

\bibitem{Kit20}
S.~Kittaka, Y.~Shimizu, T.~Sakakibara, A.~Nakamura, D.~Li, Y.~Homma, F.~Honda,
  D.~Aoki, and K.~Machida: Phys. Rev. Res. {\bfseries 2} 032014(R) (2020).

\bibitem{Cai20}
L.~P. Cairns, C.~R. Stevens, C.~D. O'Neill, and A.~Huxley: J. Phys.: Condens.
  Matter {\bfseries 32} 415602 (2020).

\bibitem{Kne19}
G.~Knebel, W.~Knafo, A.~Pourret, Q.~Niu, M.~Vali{\v{s}}ka, D.~Braithwaite,
  G.~Lapertot, J.-P. Brison, S.~Mishra, I.~Sheikin, G.~Seyfarth, D.~Aoki, and
  J.~Flouquet: J. Phys. Soc. Jpn. {\bfseries 88} 063707 (2019).

\bibitem{Ran19_HighField}
S.~Ran, I.-L. Liu, Y.~S. Eo, D.~J. Campbell, P.~Neves, W.~T. Fuhrman, S.~R.
  Saha, C.~Eckberg, H.~Kim, J.~Paglione, D.~Graf, J.~Singleton, and N.~P.
  Butch: Nature Phys. {\bfseries 15} 1250 (2019).

\bibitem{Miy19}
A.~Miyake, Y.~Shimizu, Y.~J. Sato, D.~Li, A.~Nakmura, Y.~Homma, F.~Honda,
  J.~Flouquet, M.~Tokunaga, and D.~Aoki: J. Phys. Soc. Jpn. {\bfseries 88} 063706
  (2019).

\bibitem{Kna19}
W.~Knafo, M.~Vali{\v{s}}ka, D.~Braithwaite, G.~Lapertot, G.~Knebel, A.~Pourret,
  J.-P. Brison, J.~Flouquet, and D.~Aoki: J. Phys. Soc. Jpn. {\bfseries 88} 063705 (2019).

\bibitem{Ima19}
S.~Imajo, Y.~Kohama, A.~Miyake, C.~Dong, J.~Flouquet, K.~Kindo, and D.~Aoki: J.
  Phys. Soc. Jpn. {\bfseries 88} 083705 (2019).

\bibitem{Nak19}
G.~Nakamine, S.~Kitagawa, K.~Ishida, Y.~Tokunaga, H.~Sakai, S.~Kambe,
  A.~Nakamura, Y.~Shimizu, Y.~Homma, D.~Li, F.~Honda, and D.~Aoki: J. Phys.
  Soc. Jpn. {\bfseries 88} 113703 (2019).

\bibitem{Bra19}
D.~Braithwaite, M.~Vali{\v{s}}ka, G.~Knebel, G.~Lapertot, J.~P. Brison,
  A.~Pourret, M.~E. Zhitomirsky, J.~Flouquet, F.~Honda, and D.~Aoki: Commun.
  Phys. {\bfseries 2} 147 (2019).

\bibitem{Ran20_pressure}
S.~Ran, H.~Kim, I.-L. Liu, S.~R. Saha, I.~Hayes, T.~Metz, Y.~S. Eo,
  J.~Paglione, and N.~P. Butch: Phys. Rev. B {\bfseries 101} (2020) 140503.

\bibitem{Aok20_UTe2}
D.~Aoki, F.~Honda, G.~Knebel, D.~Braithwaite, A.~Nakamura, D.~Li, Y.~Homma,
  Y.~Shimizu, Y.~J. Sato, J.-P. Brison, and J.~Flouquet: J. Phys. Soc. Jpn.
  {\bfseries 89} 053705 (2020).

\bibitem{Tho20}
S.~M. Thomas, F.~B. Santos, M.~H. Christensen, T.~Asaba, F.~Ronning, J.~D.
  Thompson, E.~D. Bauer, R.~M. Fernandes, G.~Fabbris, and P.~F.~S. Rosa: Sci.
  Adv. {\bfseries 6} eabc8709 (2020).

\bibitem{Shi20}
T.~Shishidou, H.~G. Suh, P.~M.~R. Brydon, M.~Weinert, and D.~F. Agterberg:
  arXiv:2008.04250 .

\bibitem{Ish20}
J.~Ishizuka and Y.~Yanase: arXiv:2008.01945 .

\bibitem{Kne20}
G.~Knebel, M.~Val{\v{i}}ska, M.~Kimata, F.~Honda, D.~Li, D.~Braithwaite,
  G.~Lapertot, W.~Knafo, A.~Pourret, J.-P. Brison, J.~Flouquet, and D.~Aoki: J.
  Phys. Soc. Jpn. {\bfseries 89} 053707 (2020).

\bibitem{Ike06_UTe2}
S.~Ikeda, H.~Sakai, D.~Aoki, Y.~Homma, E.~Yamamoto, A.~Nakamura, Y.~Shiokawa,
  Y.~Haga, and Y.~{\=O}nuki: J. Phys. Soc. Jpn. Suppl {\bfseries 75}
  116 (2006).

\bibitem{Har20}
H.~Harima: JPS Conf. Proc. {\bfseries 29} 011006 (2020).

\bibitem{Ish19}
J.~Ishizuka, S.~Sumita, A.~Daido, and Y.~Yanase: Phys. Rev. Lett. {\bfseries
  123} 217001 (2019).

\bibitem{Kna21}
W.~Knafo, M.~Nardone, M.~Vali{\v{s}}ka, A.~Zitouni, G.~Lapertot, D.~Aoki,
  G.~Knebel, and D.~Braithwaite: Commun. Phys. {\bfseries 4} 40 (2021).

\bibitem{Sun19}
S.~Sundar, S.~Gheidi, K.~Akintola, A.~M. C{\^o}t{\'e}, S.~R. Dunsiger, S.~Ran,
  N.~P. Butch, S.~R. Saha, J.~Paglione, and J.~E. Sonier: Phys. Rev. B
  {\bfseries 100} 140502(R) (2019).

\bibitem{Tok19}
Y.~Tokunaga, H.~Sakai, S.~Kambe, T.~Hattori, N.~Higa, G.~Nakamine, S.~Kitagawa,
  K.~Ishida, A.~Nakamura, Y.~Shimizu, Y.~Homma, D.~Li, F.~Honda, and D.~Aoki:
  J. Phys. Soc. Jpn. {\bfseries 88} 073701 (2019).

\bibitem{Kne06}
G.~Knebel, D.~Aoki, D.~Braithwaite, B.~Salce, and J.~Flouquet: Phys. Rev. B
  {\bfseries 74} 020501 (2006).

\bibitem{Par06}
T.~Park, F.~Ronning, H.~Q. Yuan, M.~B. Salamon, R.~Movshovich, J.~L. Sarrao,
  and J.~D. Thompson: Nature {\bfseries 440} 65 (2006).

\bibitem{Set11}
R.~Settai, K.~Katayama, D.~Aoki, I.~Sheikin, G.~Knebel, J.~Flouquet, and
  Y.~\={O}nuki: J .Phys. Soc. Jpn. {\bfseries 80} 094703 (2011).

\bibitem{Kim07}
N.~Kimura, Y.~Muro, and H.~Aoki: J .Phys. Soc. Jpn. {\bfseries 76} 051010 (2007).

\bibitem{Wil04}
H.~Wilhelm and D.~Jaccard: Phys. Rev. B {\bfseries 69} 214408 (2004).

\bibitem{Fis91}
R.~A. Fisher, C.~Marcenat, N.~E. Phillips, P.~Haen, F.~Lapierre, P.~Lejay,
  J.~Flouquet, and J.~Voiron: J. Low Temp. Phys. {\bfseries 84} 49 (1991).

\bibitem{Aok12_CeRu2Si2}
D.~Aoki, C.~Paulsen, H.~Kotegawa, F.~Hardy, C.~Meingast, P.~Haen, M.~Boukahil,
  W.~Knafo, E.~Ressouche, S.~Raymond, and J.~Flouquet: J. Phys. Soc. Jpn.
  {\bfseries 81} 034711 (2012).

\bibitem{Xu19}
Y.~Xu, Y.~Sheng, and Y.~feng Yang: Phys. Rev. Lett. {\bfseries 123} 217002 (2019).

\bibitem{Dua20}
C.~Duan, K.~Sasmal, M.~B. Maple, A.~Podlesnyak, J.-X. Zhu, Q.~Si, and P.~Dai:
  arXiv:2008.12377.

\bibitem{Oni00}
Y.~Onishi and K.~Miyake: J. Phys. Soc. Jpn. {\bfseries 69} 3955 (2000).

\bibitem{Wat06}
S.~Watanabe and K.~Miyake: J. Phys. Soc. Jpn. {\bfseries 75} 043710 (2006).

\bibitem{Har11}
F.~Hardy, D.~Aoki, C.~Meingast, P.~Schweiss, P.~Burger, H.~v.~Loehneysen, and
  J.~Flouquet: Phys. Rev. B {\bfseries 83} 195107 (2011).

\bibitem{Kna12}
W.~Knafo, T.~D. Matsuda, D.~Aoki, F.~Hardy, G.~W. Scheerer, G.~Ballon,
  M.~Nardone, A.~Zitouni, C.~Meingast, and J.~Flouquet: Phys. Rev. B {\bfseries
  86} 184416 (2012).

\end{thebibliography}

\end{document}